# Structured Beam Generation with a Single Metasurface

Fuyong Yue[1,⊥], Dandan Wen[1,⊥], Jingtao Xin[2], Brian Gerardot[1], Jensen Li[3], Xianzhong Chen[1*]

1. SUPA, Institute of Photonics and Quantum Sciences, School of Engineering and Physical Sciences, Heriot-Watt University, Edinburgh, EH14 4AS, UK
2. Beijing Engineering Research Centre of Optoelectronic Information and Instruments, Beijing Information Science and Technology University, Beijing, 100192, China
3. School of Physics and Astronomy, University of Birmingham, B15 2TT

**Despite a plethora of applications ranging from quantum memories to high-resolution lithography, the current technologies to generate vector vortex beams (VVBs) suffer from less efficient energy use, poor resolution, low damage threshold, bulky size and complicated experimental setup, preventing further practical applications. We propose and experimentally demonstrate an approach to generate VVBs with a single metasurface by locally tailoring phase and transverse polarization distribution. This method features the spin-orbit coupling and the superposition of the converted part with an additional phase pickup and the residual part without a phase change. By maintaining the equal components for the converted part and the residual part, the cylindrically polarized vortex beams carrying orbital angular momentum are experimentally demonstrated based on a single metasurface at subwavelength scale. The proposed approach provides unprecedented freedom in engineering the properties of optical waves with the high-efficiency light utilization and a minimal footprint.**

⊥ These authors contribute equally to this work.

* Email: x.chen@hw.ac.uk

Structured beams such as vortex beams (VBs) and vector vortex beams (VVBs) have been widely investigated as new promising resources due to the extra degree of freedom for light manipulation. VBs have a distribution of azimuthal phase and homogeneous polarization, while VVBs have inhomogeneous distribution of both phase and polarization in the transverse plane perpendicular to propagation. The applications of structured beams have been found in quantum memories[1], particle trapping[2], optical communication[3, 4], as well as high-resolution lithography[5, 6, 7, 8]. A radially polarized beam, for example, can be focused more sharply and give rise to a centred longitudinal field[6], paving the way to higher-resolution lithography and optical sensing. An azimuthally polarized beam with a helical phase front, which carries an orbital angular momentum, can effectively achieve a significantly smaller spot size in comparison with that for a radially polarized beam with a planar wavefront in a higher-NA condition[5]. What's more, the light mode with an azimuthally varying polarization and phase structure is structurally inseparable from its polarization and transverse spatial modes, known as 'classical entanglement' effect[9, 10], which has been applied in high-speed sensing[11]. Many approaches and methods, including liquid crystal q-plates[12, 13], spatial light modulator and Segnac interferometer[14], have been proposed to generate structured beams. However, these systems could not be straightforwardly downsized, preventing from widespread applications in integrated optics. In addition, the limitations of poor resolution, low damage threshold still need to be overcome for practical applications. There are numerous challenges, either fundamental or technological, in building devices that are compact, efficient and integrable.

Spin-orbit interaction describes the interaction phenomena of the polarization state and the spatial degrees of light propagating in inhomogeneous media[15, 16]. Optical metasurfaces, ultrathin inhomogeneous media with planar structures of nanopatterns that manipulate the optical properties of light at the subwavelength scale, have become a current subject of intense research due to the unprecedented control of light propagation. Metasurfaces have been widely used in many exotic research areas, including photonic spin Hall effect[17], invisibility cloaking[18], lensing[19, 20, 21, 22], and holography[23, 24]. Specially, geometric metasurfaces, regarded functionally as Pancharatnam-Berry phase optical elements[25, 26], are one of the most exciting recent advances in nano-optics due to their capability of tailoring the field of the emerging beams into nontrivial structures based on optical spin-orbit interaction[15, 16]. Geometric metasurfaces have been reported to realize orbital angular momentum generation[7, 27, 28, 29]. However, the conversion efficiency is low and only the converted part with an additional phase pickup is used. Consequently, various methods to improve the conversion efficiency[23, 30] are investigated and more optical elements are utilized to filter the residual part[27]. Nevertheless, the polarization distribution is still homogeneous. Although one possible solution is to employ two cascaded metasurfaces[31] to generate structured beam of vector vortex, the complexity of the required optical system and its large volume sharply affect its performance in system integration and competition capability. Therefore, considering the effective generation of structured light fields, an ideal solution is to use a single metasurface to generate these beams carrying an orbital angular momentum with an inhomogeneous polarization distribution.

In this work, we propose and experimentally demonstrate an approach to generate VVBs using a single reflective-type metasurface. The VVBs are generated by using the superposition of the converted part and the residual part of the output beam, which refer to the two spin eigenstates with different OAMs. The polarization states of VVBs are elucidated with a hybrid-order Poincaré sphere[32] (Hybrid-PS) whose eigenstates are defined as a fundamental-mode Gaussian beam and a Laguerre-Gaussian beam (optical vortex) with orthogonal circular polarization[33]. The designed metasurfaces to generate cylindrical VVBs carrying OAM are experimentally demonstrated and characterized. To further validate our proposed approach, a phase-gradient metasurface (PGM) is adopted as a broadband beam splitter to decompose the resultant beam into different eigenstates corresponding to the north and the south poles of the Hybrid-PS. The proposed method opens a new window to generate VVBs using a single metasurface, providing new capabilities to develop novel compact devices that may lead to advances in a wide range of fields in optics and photonics.

**Hybrid-order Poincaré sphere**

A hybrid-PS with orthogonal spin eigenstates but different values of OAM[32] is briefly described. In general, any optical wave can be decomposed into two circularly polarized helical modes carrying spin and orbital angular momentum with well defined values [34]. The light beam can be represented as

$$\left|\psi_{\ell_1,\ell_2}\right\rangle = \psi_N^{\ell_1}\left|N_{\ell_1}\right\rangle + \psi_S^{\ell_2}\left|S_{\ell_2}\right\rangle \tag{1}$$

where $\left|N_{\ell_1}\right\rangle$ and $\left|S_{\ell_2}\right\rangle$ represent two orthogonal spin eigenstates of different topological charges $\ell_1$ and $\ell_2$, can be expressed as

$$\left|N_{\ell_1}\right\rangle = \frac{1}{\sqrt{2}}\exp(i\ell_1\theta)\begin{bmatrix}1\\ \sigma i\end{bmatrix} \tag{2}$$

$$\left|S_{\ell_2}\right\rangle = \frac{1}{\sqrt{2}}\exp(i\ell_2\theta)\begin{bmatrix}1\\ -\sigma i\end{bmatrix} \tag{3}$$

Here $\exp(i\ell\theta)$ ($\ell = \ell_1$, $\ell_2$) corresponds to the azimuthal phase factor called optical vortex, whereas $\theta$ is the azimuthal angle and defined as $\theta = \arctan(y/x)$. For the general case, the integers $\ell_1$ and $\ell_2$ have different values, representing the number of the vortices' $2\pi$ helical phase windings in one wavelength. $\sigma = \pm 1$ denotes the right- and left-handed circular polarization, respectively. For $\ell_1$, $\ell_2 = 0$, the bases of Eq. (2) and (3) are reduced to the two fundamental plane wave eigenstates in a standard PS. The case $\ell_1 = -\ell_2$ corresponds to the particular case of high-order PS[35, 36]. For $|\ell_1|, |\ell_2| \geq 1$, the bases are optical vortices. $\psi_N^{\ell_1}$ and $\psi_S^{\ell_2}$ are the complex amplitudes of eigenstates, which determine the geometric position on the Hybrid-PS. The generally different $\ell_1$ and $\ell_2$ enable the most flexible description and generation of the structured beam in the current framework. In analogy, the high-order polarization state can be completely characterized in terms of the generalized Stokes parameters[35]

$$S_0 = |\psi_N^{\ell_1}|^2 + |\psi_S^{\ell_2}|^2 \tag{4}$$

$$S_1 = 2|\psi_N^{\ell_1}||\psi_S^{\ell_2}|\cos\chi \tag{5}$$

$$S_2 = 2|\psi_N^{\ell_1}||\psi_S^{\ell_2}|\sin\chi \tag{6}$$

$$S_3 = |\psi_N^{\ell_1}|^2 - |\psi_S^{\ell_2}|^2 \tag{7}$$

where $\chi = \arg(\psi_N^{\ell_1}) - \arg(\psi_S^{\ell_2})$ represents the phase difference of two eigenstates. The Hybrid-PS is then constructed with spherical coordinates using Stokes parameters. Any complete higher-order polarization state can be represented as a point on the surface of the Hybrid-PS in terms of the longitude $2\vartheta$ and latitude $2\chi$. The spherical angles $2\vartheta$ and $2\chi$ are given by

$$\tan(2\vartheta) = S_2 / S_1 \tag{8}$$

$$\sin(2\chi) = S_3 / S_0 \tag{9}$$

The north and south poles of the Hybrid-PS represent the spin-orbit eigenstates, left- and right-handed circularly polarized light with different OAMs, respectively. The hybrid-PS offers a general way to deal with high-order SAM and OAM evolution and represents VVBs. Two cases of Hybrid-PS are illustrated in Fig. 1 (a) and (b).

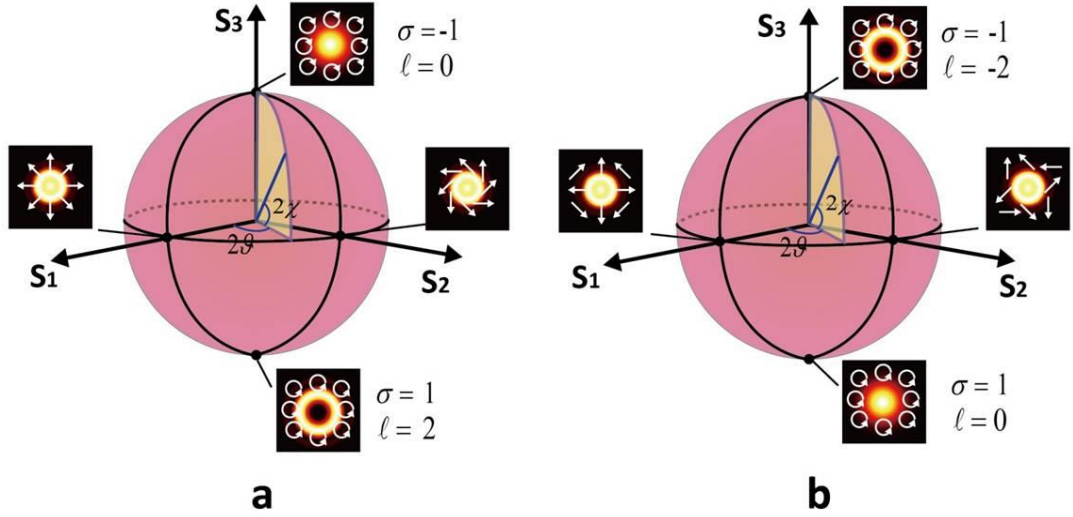

**Figure 1 | Hybrid-order Poincaré sphere with different spin-orbit eigenstates.** Any complete higher-order polarization state such as a vector vortex beam can be represented by a point on the surface of a hybrid-order Poincaré sphere. The north and south poles of the Hybrid-PS represent the spin-orbit eigenstates, left- and right-handed circularly polarized light with different OAMs, respectively. **a,** The north pole represents an eigenstate with $\sigma=-1$, $\ell=0$, and the south pole represents an eigenstate with $\sigma=1$, $\ell=2$. **b,** $\sigma=-1$, $\ell=-2$ for the north pole and $\sigma=1$, $\ell=0$ for the south pole.

**Design of the metasurface**

It is evident that any state of complete polarization on the fundamental Poincaré sphere can be described as a superposition of left- and right-handed circularly polarized light, which refer to spin eigenstates. By extending the basis of states in terms of the optical SAM to the total optical angular momentum that includes the higher-dimensional orbital angular momentum (OAM), an arbitrary optical beam with space-variant polarization and phase can be constructed from a coaxial superposition of two spin-orbit eigenstates with different values on the Hybrid-PS[37]. For example, a radially (azimuthally) polarized vortex with OAM $-\hbar$ ($\hbar$) can be generated by a superposition of two circularly polarized beams, i.e., the Laguerre-Gauss beam carrying the OAM $-2\hbar$ ($2\hbar$), and the other fundamental Gaussian beam with a plane wavefront[38]. To generate VVB, we design and fabricate reflected-type metasurfaces that simultaneously generate two eigenstates corresponding to the converted part and residual part of resultant beam with pre-designed phase difference between them. The metasurfaces have the metal-dielectric-metal configuration with the top layer of nanorods with space-variant orientation (Fig. 2 a). The high efficiency of this configuration has already been verified by numerical simulations for a uniform metasurface with all nanorod antennas aligned along the same direction (see Ref. 23 and 24).

In this context, the orientation angle of nanorods can be specified by the following expression

$$\alpha(r,\phi) = q\phi + \alpha_0 + \frac{\pi}{4} \qquad (10)$$

where $(r,\phi)$ is the polar coordinate representation, and $q$ is an integer related to the difference of topological charges for two eigenstates of Eq. (1) ($|\ell_2| - |\ell_1| = \pm 2q$). $\alpha_0$ is the initial angle related to the phase difference of two eigenstates. When a circularly polarized Gaussian beam impinges onto the metasurface, part of the incident light will be converted into the opposite circularly polarized light. The change in the SAM is transformed into OAM because of the total angular momentum conservation[27]. Therefore, the emerging beam is converted into Laguerre-Gauss beam carrying OAM with an angular index $2\sigma q$, whereas the intensity distribution of residual beam is still Gaussian with an unchanged radial index. Specially, this process is inherently geometry dependent and occurs at the subwavelength scale. By carefully designing the structure of metasurface, the converted beam and residual beam serve as two eigenstates of the Hybrid-PS. Hence, the resultant beam can be expressed as

$$E_{out}(r,\phi) = LG_{0,0}(r,\phi)\begin{bmatrix} 1 \\ \sigma i \end{bmatrix} + LG_{0,2\sigma q}(r,\phi)\exp(-i\alpha_0)\begin{bmatrix} 1 \\ -\sigma i \end{bmatrix} \qquad (11)$$

where $LG_{0,0}(r,\phi)$ and $LG_{0,2\sigma q}(r,\phi)$ represent the intensity distributions of a fundamental Gaussian beam with zero topological charge and a Laguerre-Gauss beam with topological charge of $2\sigma q$, respectively. $\exp(-i\alpha_0)$ is the phase difference between these two modes. $\sigma = \pm 1$ stands for the cases of right- and left-handed circularly polarized light, respectively.

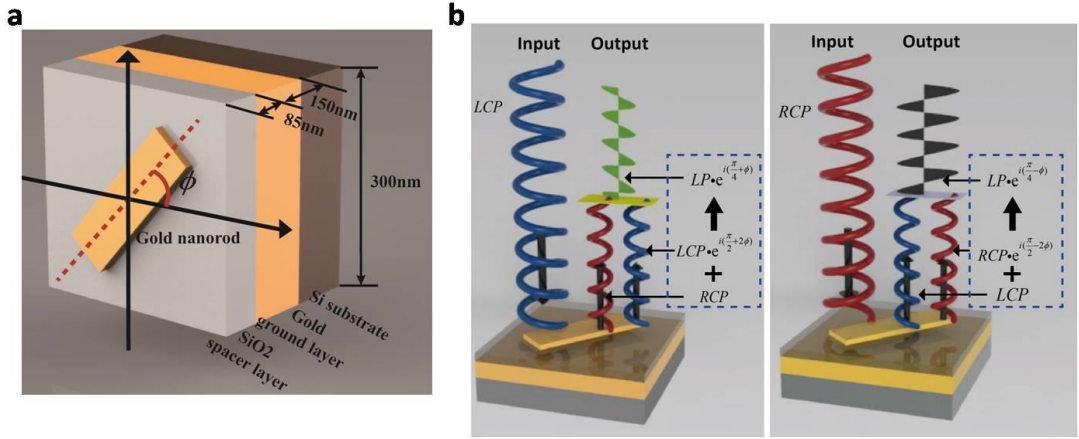

**Figure 2 | Illustration of the single-pixel cell structure and the polarization conversion of the emerging light. a,** The reflective-type half-wave plate consists of three layers: the ground gold layer (150 nm), the SiO$_2$ spacer layer (85 nm) and the top layer of gold nanorods (30 nm). Each pixel size is 300 nm by 300 nm. Each nanorod is 200 nm in length, 90 nm in width and 30 nm in thickness. **b,** For the circularly polarized incident light, the emerging light is the superposition of two orthogonal circularly polarized beams corresponding to the converted part (same handedness with the incident beam) and residual part (opposite handedness with the incident beam), respectively. When the converted part and the residual part have equal components, the output beam gives rise to

the linearly polarized light. Spiral curves in blue colour stands for the left-handed circularly polarized light (LCP), and that in red colour stands for the right-handed circularly polarized light (RCP).

To clearly illustrate the mechanism, the metasurface is illuminated at normal incidence by a right-handed circularly polarized light (RCP) beam with a normalized Jones vector $\frac{1}{\sqrt{2}}\begin{bmatrix}1\\-i\end{bmatrix}$. It is worth mentioning that the handedness of polarized light is reversed when it is reflected by an ideal mirror because of the opposite propagation direction. At a particular location of azimuthal angle $\phi$, the polarization state of the resultant beam is generally a superposition of two components with orthogonal circular polarization states, i.e., the converted part with an abrupt phase change $E_{Con}^{RCP}$ and the residual part without phase delay $E_{Res}^{RCP}$. Considering $q=1$ and $\alpha_0=0$, the expression for the superposition is given by

$$E_{out}^{RCP} = (E_{Con}^{RCP} + E_{Res}^{RCP}) = \frac{1}{\sqrt{2}}(\frac{A}{A+B}e^{i(\frac{\pi}{2}-2\phi)}\begin{bmatrix}1\\-i\end{bmatrix} + \frac{B}{A+B}\begin{bmatrix}1\\i\end{bmatrix}) \qquad (12)$$

Apart from the Ohm loss and absorption, all of the reflected beams contribute to VVB generation. *A* and *B* represent the amplitudes of converted and residual light, respectively. If we tune the reflection properties of the gold nanorods to have $A=B$, the resultant beam gives rise to the linearly polarized light and also acquires a phase change. Its Jones vector is given by

$$E_{out}^{RCP} = \sqrt{2}e^{i(\frac{\pi}{4}-\phi)}\begin{bmatrix}\cos(\frac{\pi}{4}-\phi)\\ \sin(\frac{\pi}{4}-\phi)\end{bmatrix} \qquad (13)$$

A similar derivation takes place when the polarization state of the incident light is LCP. The Jones vector of emerging beam is

$$E_{out}^{LCP} = \sqrt{2}e^{i(\frac{\pi}{4}+\phi)}\begin{bmatrix}\cos(\frac{\pi}{4}+\phi)\\ -\sin(\frac{\pi}{4}+\phi)\end{bmatrix} \qquad (14)$$

The superscripts in Eqs. (12-14) represent the polarization states of the incident light. Figure 2 b shows the polarization and the phase evolution when a circularly polarized incident light passing through a metasurface. If $A \neq B$, the resultant beam is elliptically polarized with an additional phase change. In brief, our metasurface is able to generate a structured beam from an unstructured one by using the spin-orbit coupling as a method of entanglement. As a result, there is also a global phase factor, in the above case $\exp(i(\phi+\pi/4))$, appears together with the polarization distribution.

Figure 3 shows the polarization and phase distribution of the input and output beam when reflected from the metasurface with $q=1$ and $\alpha_0=0$. For an incident LCP Gaussian beam with zero OAM, part of the input beam is converted into the Laguerre-Gauss beam carrying OAM of $2\hbar$ due to the spin-to-orbit coupling[27]. In contrast, the residual beam is still a Gaussian beam but has the opposite handedness in comparison with that for the converted part. The output beam is therefore a superposition of two components, and thus the polarization distribution is transformed to a radial vector field when these two components have equal intensity according to Eq. (14) (top in Fig. 3). More importantly, the OAM is down to $\hbar$. Furthermore, the polarization distribution is flipped about the vertical axis when switching the incident light from LCP to RCP according to Eq. (13) (bottom in Fig. 3).

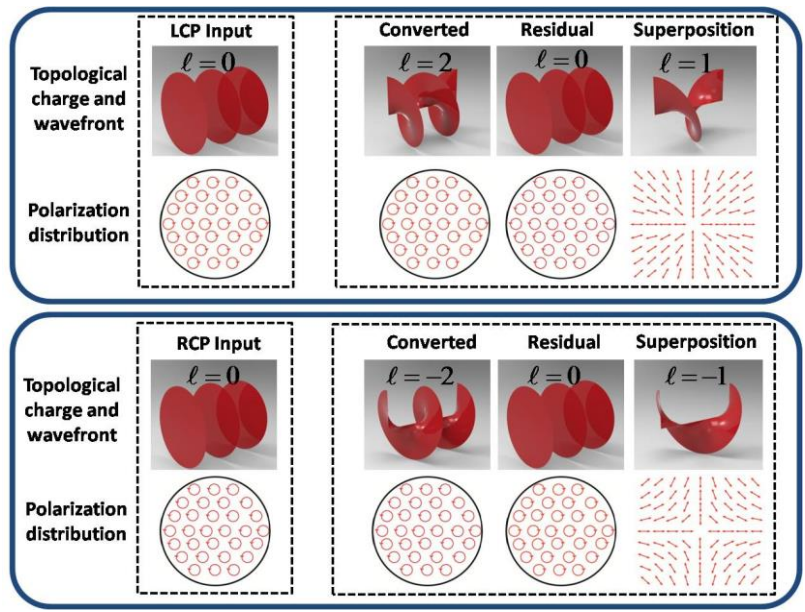

**Figure 3 | Illustration of topological charge of orbital angular momentum, wavefront and polarization distribution of incident light, converted, residual and resultant beams.** The resultant beam is the superposition of the converted and residual part from the metasurface with $q=1$ and $\alpha_0=0$. $q$ is an integer related to the difference of topological charges for two eigenstates($|\ell_2|-|\ell_1|=\pm 2q$). $\alpha_0$ is the initial angle related to the phase difference of two eigenstates. The upper figure and lower show the cases for LCP and RCP incident light, respectively.

**Results**

To verify the proposed approach and explore the device performance, we design and fabricate two metasurfaces with $q=1$ and different values of $\alpha_0$. The gold ground layer and SiO$_2$ spacer layer are deposited on a silicon substrate by electron beam evaporation (see Fig. 2). The top layer of nanorods is fabricated using electron-beam lithography and standard lift-off process, and a 3-nm-Ti layer is deposited between the top layer and the SiO$_2$ spacer layer for adhesion purpose. In order to verify our proposed method and measure their respective intensities of the two eigenstates, we employ a phase-gradient metasurface (PGM), which is used as a special beam splitter in experimental setup that allow us, without the need of any additional polarizer and waveplate, to simultaneously decompose the output beam from the metasurface into two eigenstates on the Hybrid-PS. The schematic of experimental setup is depicted in Fig.4. Here, the metasurface sample for the VVB generation is mounted on a three-dimensional translation stage and exposed to the light from tunable NKT supercontinuum laser. A Glan polarizer (GP) and a quarter-wave plate (QWP) are used to generate the required circularly polarized light. Then the light is weakly focused by a lens with a focal length of 100 mm onto the metasurface to ensure that the beam size is smaller than the sample. In order to collect the reflected light, a polarization-insensitive beam-splitter (BS) is inserted between the QWP and the lens. The reflected vector vortex beam is either projected to PGM to decompose the resultant beam into two eigenstates, or propagates in the free space for further application. The SEM images of the fabricated metasurface for VVB generation and PGM for eigenstates decomposing are shown in Fig.5 a and b, respectively. Fig. 5 c shows the measured normalized power of the two eigenstates over a wide range of wavelengths. The two curves overlap at the wavelength of 697 nm, which means that the two eigenstates have equal components at this wavelength and the VVB is realized. Importantly, the power ratio of eigenstates determines the location of the higher-order states on Hybrid-PS. While the mentioned radially polarized beam is optimally generated at this wavelength, by changing the wavelength of incident light, other high-order polarization states on Hybrid-PS can also be realized as well. The simulated and obtained intensity distributions of two eigenstates are shown in Fig. 5 d. The doughnut shape and singular point confirm the existence of optical vortex (right in Fig. 5 d), which corresponds to the converted light. The residual part (left in Fig. 5 d), on the other hand, is confirmed by the shape without a singularity in the light spot corresponding to the Gaussian beam. The middle image in Fig. 5 d is the profile of the original vector vortex beam resultant from our metasurface in generating structured light, which is the superposition of the optical vortex and the Gaussian beam with equal intensities. The experimental results are in good agreement with the simulation results. To further reveal the spiral wavefront and verify the OAM of optical vortex, the converted beam is interfered with the co-propagating Gaussian beam[39]. The double helical intensity profile and the number of branches stemming from the singularity confirm that the converted beam carries orbital angular momentum of $2\hbar$ (Fig. 5e). Switching the circular polarization of incident light from LCP to RCP introduces the change of the twisted direction of the helical wavefront (displayed on the interference pattern) and the sign of OAM value (from '+' to '-').

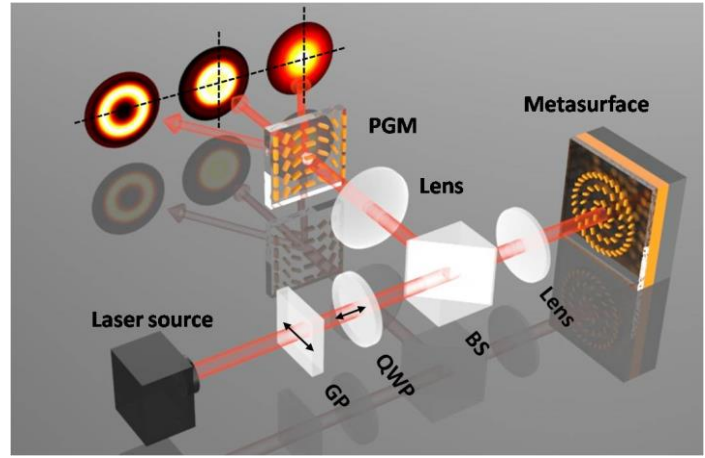

**Figure 4 | Schematic of the experimental setup for eigenstate characterization.** The required circularly polarized light beam is generated by a laser beam (NKT-SuperK EXTREME) passing through a Glan polarizer (GP) and a quarter-wave plate (QWP), then impinges upon metasurface at normal incidence with a weak focus by a lens with a focal length of 100 mm. In order to collect the reflected light and project it to a phase-gradient metasurface (PGM), a polarization insensitive beam splitter (BS) is inserted between QWP and lens. A charge coupled device (CCD) camera is used to image the output beams.

The resultant beam is transformed to vectorial beam with OAM $-\hbar$ ($\hbar$) when two eigenstates have equal intensities upon the illumination of RCP (LCP) light, which can be discribed by a point lying on the equator of Hybrid-PS (see Fig. 1 **a** and **b**). We numerically and experimentally characterize the vectorial vortex carrying OAM by using an analyzing polarizer at different polarization angles.

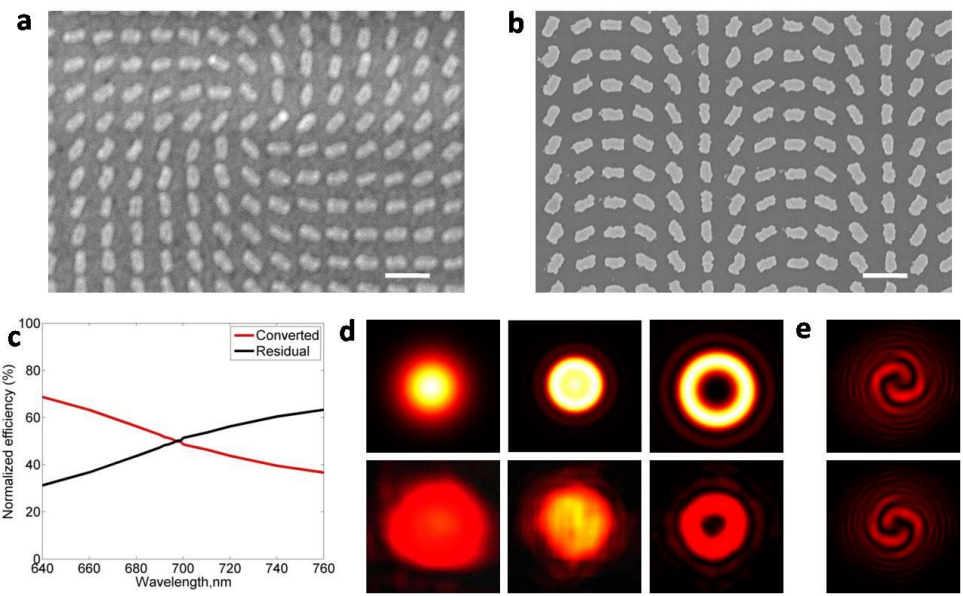

**Figure 5 | Characterization of two eigenstates generated by a metasurface with *q*=1 and $\alpha_0 = 0$. a**, Scanning electron microscopy image of the metasurface for vector vortex beam generation. **b**, Scanning electron microscopy image of the phase gradient metasurface. **c,** Measured power (normalized) of two eigenstates at various wavelengths. **d**, Theoretically predicted and measured intensity distribution of the two eigenstates (left and right side) and original beam profile before decomposition (middle). **e**, Spiral patterns created by the interference of the vortex beam and a co-propagating Gaussian beam. The polarization states of incident light are LCP (upper image) and RCP (lower image), respectively.

By passing through a rotating polarizer[31], the generated structured beams from the fabricated metasurfaces are characterized and validated. Figure 6 shows the simulated and measured intensity distribution of vectorial vortex after passing through an analyzing polarizer in horizontal, 45°, vertical, and − 45 ° orientations at the wavelength of 697 nm. The appearance of '*s*' shape patterns is theoretically predicted and experimentally confirmed. The observed patterns indicate that the resultant beams indeed have an inhomogeneous polarization distribution and a helical wavefront. Moreover, the twisted direction of the '*s*' shape varying with the handedness of circular polarization are also experimentally confirmed from the obtained intensity patterns.

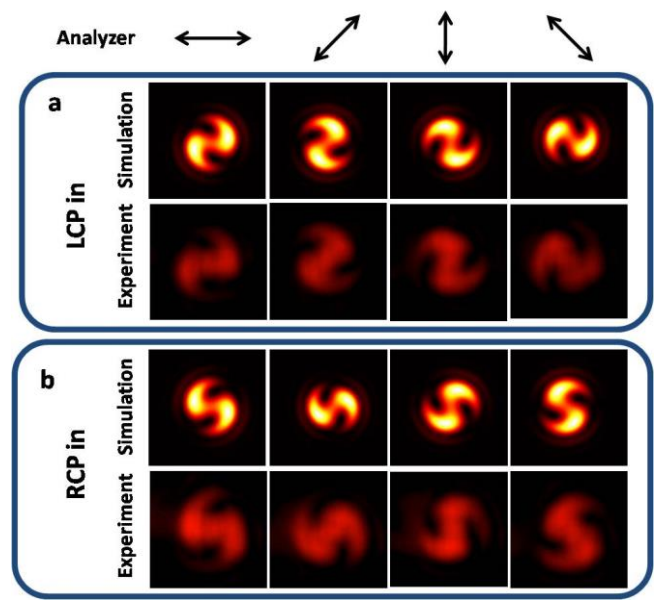

**Figure 6 | Simulated and experimentally recorded intensity distribution of the vector vortex after passing through a polarizer with different polarization angles.** The polarization angles of the polarizer include horizontal, diagonal, vertical and antidiagonal directions. The top row shows the different polarization angles of the polarizer. The polarization states of the incident light are a) LCP and b) RCP, respectively.

**Discussion**

Efficient generation of the structured optical fields using an ultra-compact device has both fundamental and technical importance for photonics related research. As promising candidates for integrated optics, metasurfaces have opened a broad range of applications for inhomogeneous control amplitude, phase, and polarization of the scattered waves. The proposed method provides an unusual way to generate vector vortex beam carrying orbital angular momentum using a single metasurface, which will inspire the pursuit of further novel functionalities. To our best knowledge, this is the first time that the converted part and the residual part are used together to realize new functionalities to achieve the most efficient light utilization. Both the phase and polarization are manipulated at subwavelength scale by the artificial array of engineered nanorods over the metasurface. On the other hand, both the polarization distribution and the orbital angular momentum of the output wave are controlled by the handedness of the incident light polarization. For Hybrid-PS polarization generation, two eigenmodes are generated simultaneously and exactly co-axis. This design significantly simplifies the experimental setup and shrinks the size of the photonic device, paving the way for the operation of compact integrated devices. We develop two metasurfaces with same $q=1$ but different initial angles ($\alpha_0 = 0$ and $\alpha_0 = \pi$), which can generate radially and azimuthally polarized vortices carrying orbital angular momenta, respectively. The two types of structured beams can be geometriclly represented as two spots on the equator of the Hybird-PS. Figure 7 shows the the intensity patterns for the metasurface with $q=1$ and $\alpha_0 = 0$ and that with $q=1$ and $\alpha_0 = \pi$. The SEM image of the metasurface with $\alpha_0 = \pi$ is available in Fig. S5. The good agreement between predicted and experimental results confirms the proposed methodology. It has been reported that azimuthally polarized beams with helical wavefront could effectively achieve a significantly smaller spot than normal azimuthally polarized beams when focused with a high-numerical-aperture lens[5].

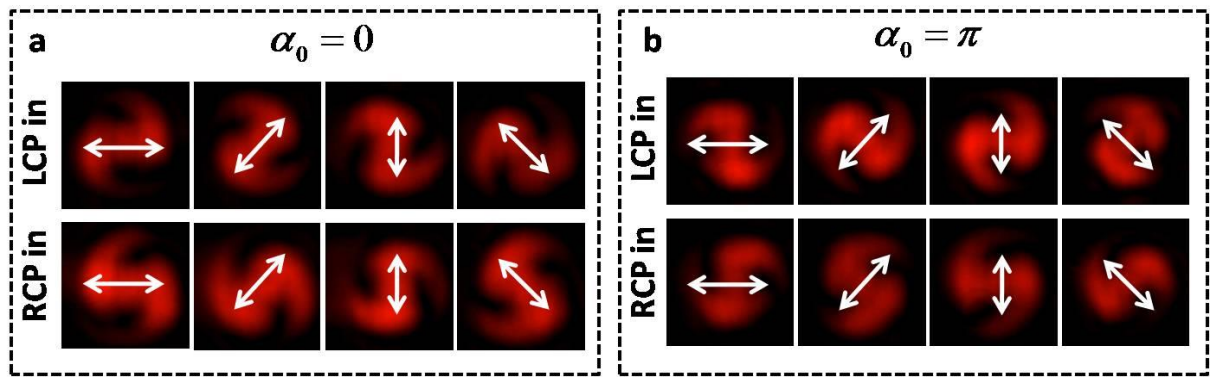

**Figure 7 | Measured intensity distribution of vector vortex beams generated by metasurfaces with *q*=1** ($\alpha_0 = 0$ **and** $\alpha_0 = \pi$) **after passing through an analyzing polarizer with different orientations. a,** The intensity distributions of metasurface with *q*=1 and $\alpha_0 = 0$. The top row is for the case of LCP input light, and the bottom row is for the case of RCP input light. White arrows show the orientation of polarizer. **b,** The intensity distributions of metasurface with *q*=1 and $\alpha_0 = \pi$.

For the application of free space communication, the optical vortex has attracted growing attention due to its higher data transmission capacity. However, the atmospheric turbulence strongly affects the properties of the optical vortex when propagating in free space. The existence of the vectorial vortex can be identified with longer propagation distance through atmosphere than the scalar vortex even with vanishing characteristic vortex structure[40]. What's more, if the angles of the nanorods are the same, then the Hybrid-PS will be reduced to a fundamental Poincaré sphere. By carefully designing the angle distribution of nanorods, any polarization state can be realized using a single metasurface with the circularly polarized incident light. In addition, the angle of resultant polarization is switchable by controlling the handedness of the circularly polarized incident light.

In summary, we propose and experimentally demonstrate an approach to generate vector vortex beams using a single plasmonic metasurface. The uniqueness of the proposed method lies in the combined action of the Pancharatnam-Berry phase and the superposition of two orthogonal circularly polarized light beams. Both orbital angular momentum and polarization distribution in transverse plane about propagation axis are manipulated by a single metasurface consisting of nanorods with spatially variant orientation. As our work solves several major issues typically associated with VVB generation: less efficient energy use, poor resolution[12], low damage threshold[13], bulky size and complicated experimental setup[31], it opens a new window for future practical applications of the structured beams in the relevant research fields such as optical communication[3], particle trapping[2], microscopy and quantum optics[1].